\documentstyle[pramana,floats,psfig]{ias}
\begin{document}
\mark{{Spinodal decomposition...}{P. Shukla and A. K. Mohanty}}
\title{Spinodal decomposition: An alternate mechanism of 
       phase conversion}

\author{P. Shukla and A. K. Mohanty}
\address{Nuclear Physics Division,
Bhabha Atomic Research Centre,\\
Trombay, Mumbai 400 085, India}
\abstract
{ The scenario of homogeneous nucleation is investigated for a first 
order quark-hadron phase transition in a rapidly expanding background of
quark gluon plasma. It is found that significant supercooling is 
possible before hadronization begins. This study also suggests that 
spinodal decomposition competes with nucleation and may provide
an alternative mechanism for phase conversion.}

\maketitle

\section{Introduction}
If quark gluon plasma (QGP) is formed in relativistic heavy ion collisions
then a phase transition from QGP to normal hadron matter must take place 
during some time of its expansion. 
Below critical temperature $T_C$ (supercooled), the QGP phase becomes 
metastable and phase transition can be initiated by the 
nucleation of critical size hadron bubbles in a homogeneous QGP phase.
The amount of supercooling
depends on the strength of the transition being more for a strong 
first order transition. As the system supercools, the barrier separating 
the metastable QGP phase from the stable hadron phase 
decreases and the system may reach the point of spinodal instability $T_S$, 
at which the barrier vanishes. Hence, for a strong enough supercooling, 
the spinodal decomposition may provide an alternate path for phase conversion. 
In case of homogeneous nucleation, the system requires to cool upto 
a temperature $T_m$ at which appreciable nucleation is reached and 
system will be reheated due to the release of latent heat. 
But if the temperature $T_m$ is less than $T_S$ 
then the phase transition may proceed via spinodal
decomposition.

 For the case of QGP produced in heavy ion collision,
the route to hadronization either through nucleation or
through spinodal decomposition
depends sensitively on several factors like nucleation rate
and expansion scenario.
 In this work we study amount of supercooling and
rate of hadronization by solving self consistently the nucleation rate along
with hydrodynamic equation corresponding to 
both longitudinal and spherical expansion.

\section{Spinodal and supercooling temperature}

As the order as well as the strength  of the quark hadron phase transition
is still an unsettled issue, we consider a more generic form of the
potential which covers a wide range from very strong to very weak first
order phase transition \cite{NHOMO}
\begin{eqnarray}\label{vt}
V(\phi, T) &=& a(T)\,\phi^2 - b \,T \,\phi^{3 } +  c \, \phi^4,
\end{eqnarray}
where $b$ and $c$ are  positive constants which can be
expressed in terms of surface tension $\sigma$ and correlation
length $\xi$.
 The requirement that at all the temperatures,
the difference between the two minima
should be equal to the pressure difference between
the two phases fixes the third parameter $a(T)$.
  The spinodal temperature $T_S$ where the quark phase becomes
unstable and there exists only one minimum corresponding to
hadron phase, can be obtained as \cite{SPINO}
\begin{eqnarray}\label {tspin}
T_S = \left[\frac{B}{B + 81\sigma/16\xi} \right]^{1/4} \, T_C,
\end{eqnarray}
 where $B$ is the bag constant.
If the QGP supercools upto this point, 
it will become unstable and may go to hadron phase by spinodal 
decomposition.
 For a strong enough transition, $\sigma/\xi$ is large
and $T_S$ is lower as compared to the case when the transition is weak.
We are interested to know whether the system cools down to the
temperature $T_S$. For comparison, we denote the minimum temperature
reached during the supercooling as $T_m$. 
Both $T_S$ and $T_m$ depend on the strength of the transition and need
to be evaluated properly. While $T_S$ can be estimated directly
from Eq.~(\ref{tspin}), $T_m$ requires a self consistent solution of a
set of equations involving
the nucleation rate and energy momentum conserving hydrodynamical equations
as described in Ref.~\cite{SPINO}

The evolution of the energy density is given by  \cite{5DAN}
\begin{eqnarray}\label{hydro}
\frac{d e}{d\tau} + \frac{D \,\omega}{\tau} = {4\eta/3 + \zeta \over \tau^2},
\end{eqnarray}
where $D$=1 for the expansion in $(1+1)$ dimension.
The factors $\eta$ and $\zeta$
are the shear and the bulk viscosity of the medium.
For nonviscous plasma (for zero viscosity), the above equation
follows the Bjorken's scaling solution where $T^3\tau$ is a constant.
The energy momentum
equation needs to be solved numerically for expansion in $(3+1)$
dimensions. However, retaining the simplicity,
we can still use Eq.~(\ref{hydro}) for spherical expansion \cite{5KAPZ},
with the choice of $D=3$.

\section{Results and Discussions}

The critical temperature is fixed at $T_C$ = 160 MeV.
The correlation length is fixed at $\xi=0.7$ fm and
$\sigma$ is treated  as a free parameter in the present study.
The time $\tau_C$ that the plasma takes to cool down to $T=T_C$
will depend on the initial conditions and on the expansion dynamics.
We assume it of the order of  4 to 8 fm/$c$.
From $\tau_C$ onwards, we consider longitudinal expansion
if it is 4 fm/$c$ and spherical expansion if it is
between 6 to 8 fm/$c$.

\begin{figure}
\centerline{\hbox{
\psfig{figure=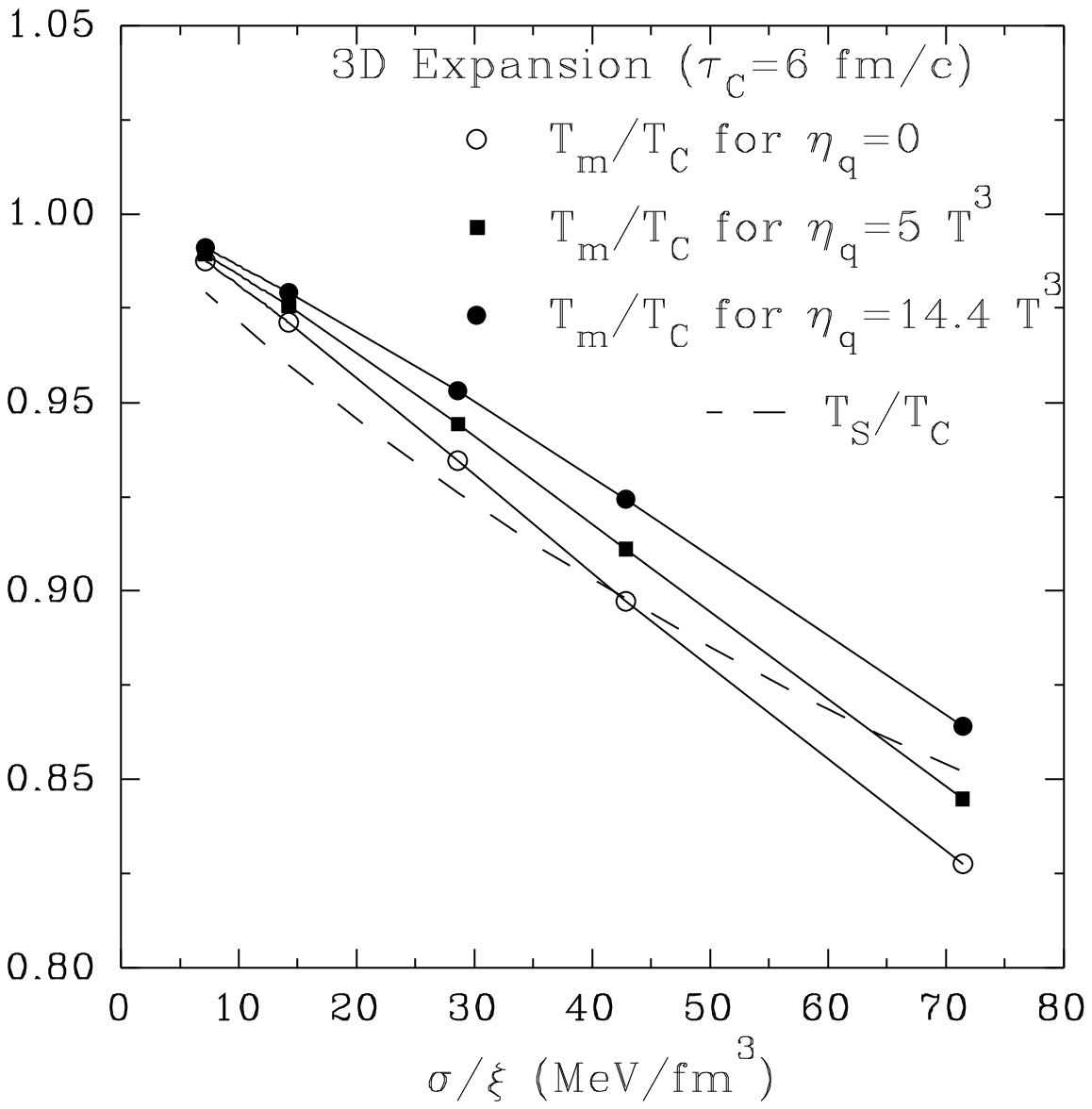,height=6cm,width=6cm} 
\psfig{figure=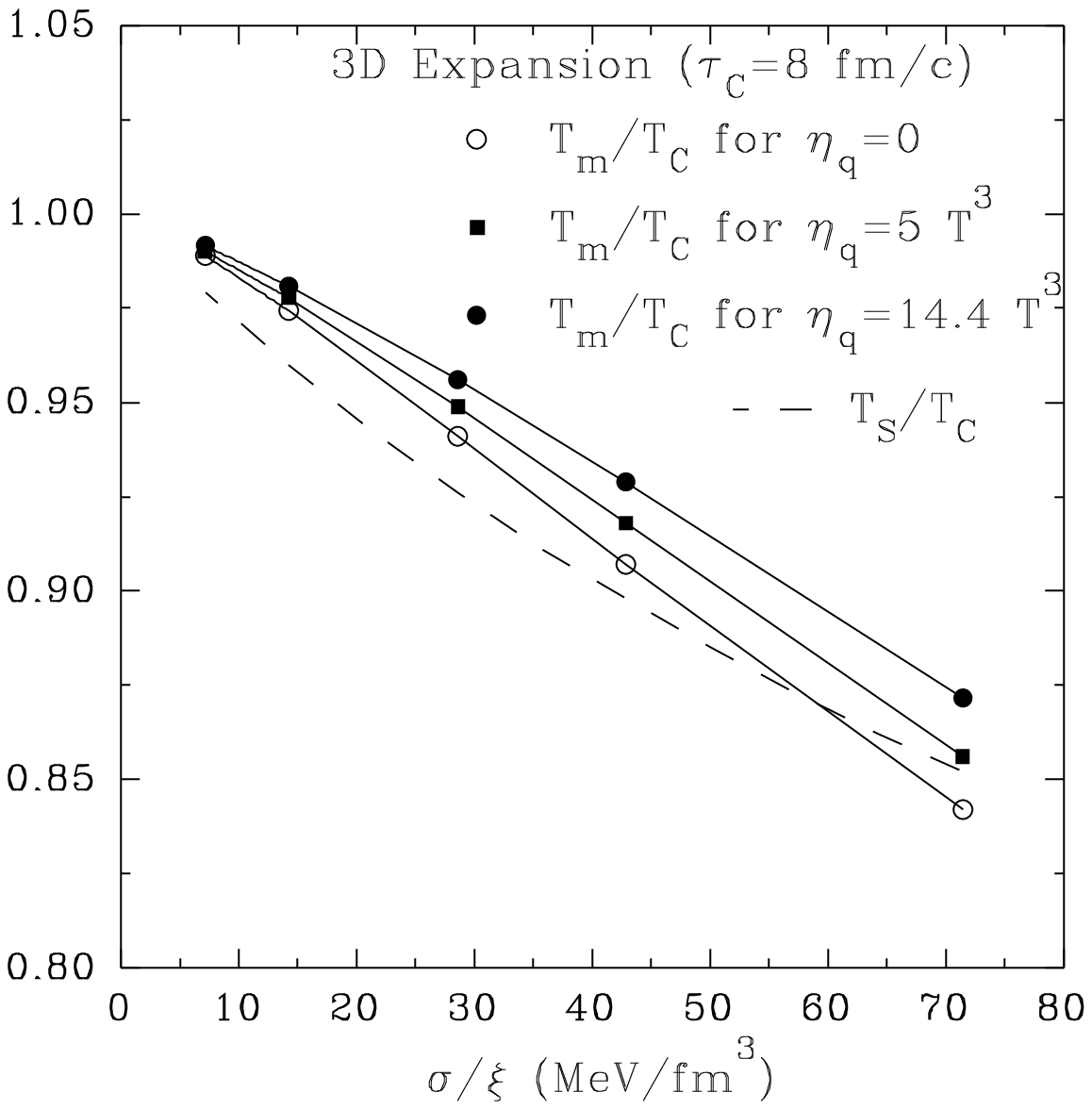,height=6cm,width=6cm}}}
\caption{The spinodal temperature $T_S/T_C$ and minimum
temperature $T_m/T_C$ reached during supercooling}
\label{f12}
\end{figure}

\begin{figure}
\centerline{\hbox{
\psfig{figure=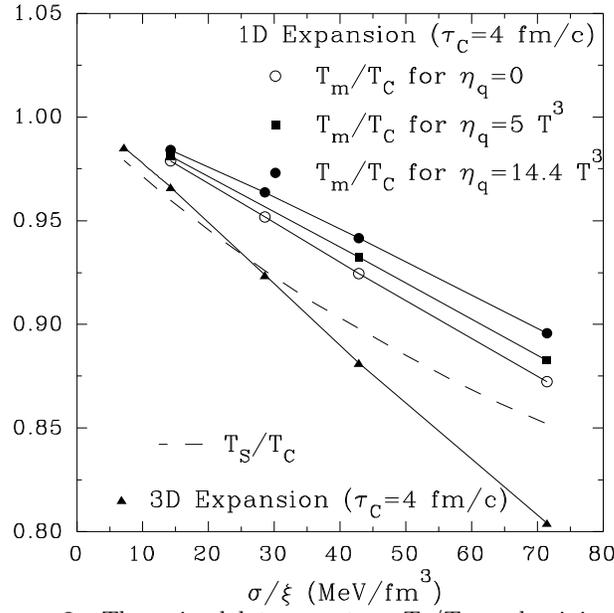,width=8cm,height=8cm}}}
\caption{The spinodal temperature $T_S/T_C$ and minimum
temperature $T_m/T_C$ reached during supercooling}
\label{fig61}
\end{figure}

 Figure~(\ref{f12}) shows a plot of $T_m/T_C$ and $T_S/T_C$ as a function 
of strength of the transition, $\sigma/\xi$ for different values of 
viscosity coefficient $\eta_q$ at $\tau_C$=$6$ fm/$c$ and $8$ fm/$c$ 
respectively in left and right panels.
 First consider the case of nondissipative plasma with viscosity
$\eta_q=0$. The curves $T_m$ and $T_S$ show
a cross over point as the transition becomes 
stronger (large $\sigma/\xi$). For weak enough transition, $T_m$ is
well above the spinodal temperature $T_S$. Since $T_m$ depends on
expansion rate of the medium, the cross over point will sensitively
depend on $\tau_C$; moving towards right for the slower 
expansion (higher $\tau_C$).
If the plasma is viscous [non-zero $\eta_q$] the 
$T_m$ is increased further due to slow evolution of the medium.
Even the cross over point also shifts towards right showing that the 
nucleation is the dominant mechanism over a wide range 
of $\sigma/\xi$ ratios.
  Figure~(\ref{fig61}) shows a similar plot for a longitudinally 
expanding system. For comparison the calculation for spherical
expansion is also shown which gives the lower limit on $T_m$.
One can see that for longitudinal expansion, the system remains
far from any spinodal instability.

 Similar work has been carried out by Scavenious 
et al. \cite{5DUMHEP}, using linear sigma model. 
We use bag model equation of state throughout
our work and also incorporate a more relevant expansion scenario.

\section{Conclusions}
 Although the above study depends on the choice of the parameters, 
a general observation is that for strong enough 
transition (large $\sigma/\xi$) with zero or very small amount
of viscosity, $T_m < T_S$; The system reaches the spinodal instability 
before the amount of nucleated hadron bubbles become significant to begin 
phase conversion. The phase conversion in such a case
will proceed via spinodal decomposition.
If the medium is viscous or the transition is weak enough or both,
the supercooling is much less and $T_m > T_S$; The phase conversion 
may still proceed through
homogeneous nucleation. However, depending on the range of the parameters,
there could be a competition between the homogeneous nucleation and the
spinodal decomposition as the nucleation and expansion
time scales for relativistic heavy ion collisions are comparable.


\begin{thebibliography}{99}
\bibitem{NHOMO}  P. Shukla, A. K. Mohanty, S. K. Gupta, and M. Gleiser,
       Phys. Rev. C{\bf 62}, 054904 (2000).

\bibitem{SPINO} P. Shukla and A. K. Mohanty, 
       Phys. Rev. C{\bf 64}, 054910 (2001).

\bibitem{5DAN} P. Danielewicz and M. Gyulassy, Phys. Rev. D{\bf31}, 53 (1985).

\bibitem{5KAPZ} L.P. Csernai and J.I. Kapusta, Gy. Kluge, and
            E.E. Zabrodin, Z. Phys. C{\bf 58}, 453 (1993).

\bibitem{5DUMHEP} O. Scavenious, A. Dumitru, E.S. Fraga, J.T. Lenaghan,
         A.D. Jackson, Phys. Rev. D{\bf 63}, 116003 (2001).
\end{thebibliography}
\end{document}